\documentclass[conference,9pt]{IEEEtran}
\IEEEoverridecommandlockouts

\usepackage{cite}
\usepackage{blindtext, graphicx}
\usepackage[cmex10]{amsmath}
\usepackage{amsfonts,amsthm,amssymb}
\usepackage[font=scriptsize,caption=false,labelsep=space]{subfig}
\usepackage{graphicx,float,wrapfig}
\usepackage{graphicx}
\usepackage{graphics,color}
\usepackage{mathtools}
\usepackage{nicefrac}
\usepackage{bm}
\usepackage{txfonts}

\def\IEEEbibitemsep{0pt plus .5pt}
\usepackage{balance}

\usepackage{color}

\usepackage{wrapfig} 

\usepackage{upquote}
\usepackage[none]{hyphenat}

\begin{document}
\setlength{\abovedisplayskip}{3pt}
\setlength{\belowdisplayskip}{3pt}
%
\title{Sub-Nyquist Channel Estimation over IEEE 802.11ad Link\vspace{-28pt}}

\author{\IEEEauthorblockN{Kumar Vijay Mishra and Yonina C. Eldar}\thanks{The authors are with the Andrew and Erna Viterbi Faculty of Electrical Engineering, Technion - Israel Institute of Technology, Haifa, Israel. E-mail: \{mishra, yonina\}@ee.technion.ac.il.
This project is funded by the European Union's Horizon 2020 research and innovation programme under grant agreement ERC-BNYQ. K.V.M. acknowledges partial support via Andrew and Erna Finci Viterbi Fellowship and Lady Davis Fellowship.}
}

\maketitle

\begin{abstract}
Nowadays, millimeter-wave communication centered at the 60 GHz radio frequency band is increasingly the preferred technology for near-field communication since it provides transmission bandwidth that is several GHz wide. The IEEE 802.11ad standard has been developed for commercial wireless local area networks in the 60 GHz transmission environment. Receivers designed to process IEEE 802.11ad waveforms employ very high rate analog-to-digital converters, and therefore, reducing the receiver sampling rate can be useful. In this work, we study the problem of low-rate channel estimation over the IEEE 802.11ad 60 GHz communication link by harnessing sparsity in the channel impulse response. In particular, we focus on single carrier modulation and exploit the special structure of the 802.11ad waveform embedded in the channel estimation field of its single carrier physical layer frame. We examine various sub-Nyquist sampling methods for this problem and recover the channel using compressed sensing techniques. Our numerical experiments show feasibility of our procedures up to one-seventh of the Nyquist rates with minimal performance deterioration.
\end{abstract}
\vspace{-6pt}
\vspace{-10pt}
\section{Introduction}
\label{sec:intro}
The pervasive presence of wireless networks in indoor data communication has led to the demand for very high data throughput links. In this context, recent advances in millimeter-wave technology provide multi-gigabit connectivity in the 60 GHz unlicensed radio-frequency (RF) band with 7-9 GHz of contiguous spectrum \cite{rappaport2013millimeter}. The IEEE 802.11ad standard has been recently developed to enable throughput of up to 7 Gbps at 60 GHz for short-range wireless communication \cite{ieee2012phy80211ad}. The diminished range is due to heavy attenuation through physical barriers at 60 GHz. However, in an indoor wireless setting, such a high-speed link is ideal to enable applications which require huge data rates such as transmission of uncompressed high definition (HD) video \cite{daniels200760}, in-room gaming \cite{agilent2013wireless}, intra-large-vehicle communications \cite{choi2016millimeter,gonzalez2016radar} and indoor positioning systems \cite{chen2016indoor}. 

The IEEE 802.11ad transmits both single carrier (SC) and orthogonal frequency-division multiplexing (OFDM) modulation schemes at operating frequencies of 1.76 GHz and 2.64 GHz, respectively. The SC physical layer (SCPHY) frame of 802.11ad protocol encapsulates Golay complementary sequences \cite{golay1961complementary,golay1962note}. These sequences have the property of perfect \textit{aperiodic} autocorrelation (zero sidelobes) making them useful for communication channel estimation \cite{mo2016channel,qureshi2009mimo} and radar remote sensing \cite{farnett1990pulse,chi2011complementary}. In this paper, we focus on the channel estimation problem using 802.11ad SCPHY frames.

Environments such as indoor wireless networks and vehicular communications are highly variable with typical channel coherence times of nanoseconds. As a result, 802.11ad chip sequences are of extremely short duration requiring the receivers to use analog-to-digital converters (ADCs) that operate at very high ($\sim$GHz) sampling rates. It is desirable to avoid these high-rate ADCs for the purposes of saving power, circuit area and cost. In this context, several methods have been proposed recently to address the problem of channel estimation for ultra wideband wireless systems using fewer samples than obtained at the Nyquist rate (see \cite{cohen2014channel} and references therein). Most exploit the fact that the channel is \textit{sparse} facilitating the use of compressed sensing (CS) techniques \cite{eldar2012cs,eldar2015sampling}. 

However, sub-Nyquist channel estimation for IEEE 802.11ad remains relatively unexamined in prior studies. In this paper, we develop an 802.11ad-based channel estimation procedure that operates at up to one-seventh of the Nyquist rate without significant loss of performance. Although a comprehensive statistical characterization of the 802.11ad channel model is still ongoing, it has been found that it is similar to the well-documented IEEE 802.15.3c channel model \cite{rappaport2014millimeter,suryaprakash2016millimeter}, and can be assumed to be sparse. Our recovery algorithm uses the \textit{Xampling} framework where Fourier coefficients of the received signal are acquired from their low-rate samples \cite{eldar2015sampling}.

Among prior studies, the closest to our work is \cite{cohen2014channel} where analog sub-Nyquist sampling methods are developed for channel estimation. The transmit waveform in \cite{cohen2014channel} is an Ipatov sequence that has the property of perfect \textit{periodic} autocorrelation. For the IEEE 802.11ad link, \cite{liu2015all,ye2013improved} explore channel estimation at Nyquist sampling rates. Recently, the special structure of 802.11ad SCPHY frame has also been exploited for parameter estimation in an automotive radar setup that operates at Nyquist rates and uses Wireless Local Area Network (WLAN) for vehicle-to-vehicle communications \cite{kumari2015investigating}. CS-based channel estimation for 802.11ad link is investigated in \cite{suryaprakash2016millimeter} where it is assumed that low rate measurements are already available. Compared to these studies, our method directly deals with the sub-Nyquist sampling of the IEEE 802.11ad waveform, leverages the special construction of the IEEE 802.11ad SCPHY frame, and suggests feasible modifications in 802.11ad hardware for possible implementations. As in \cite{cohen2014channel}, we observe that foldable sampling leads to the best performance over all hardware realizable sub-Nyquist methods examined.

In the next section, we describe the special structure of the IEEE 802.11ad SCPHY preamble including properties of Golay complementary sequence pairs. Then, we explain our system model and conventional Nyquist-based channel estimation. We present sub-Nyquist channel estimation in Section IV. Various sampling strategies to obtain few Fourier coefficients are discussed in Section V where we validate our model and methods via numerical experiments. We provide concluding remarks in Section VI.
\vspace{-9pt}
\section{System Model and Nyquist Processing}
\label{sec:nyq}
The IEEE 802.11ad protocol frame has a complex structure. We rely only on its SCPHY preamble for channel estimation.
\vspace{-8pt}
\subsection{Transmitter}
\label{subsec:nyqtx}
\begin{figure}
  \includegraphics[scale=0.25]{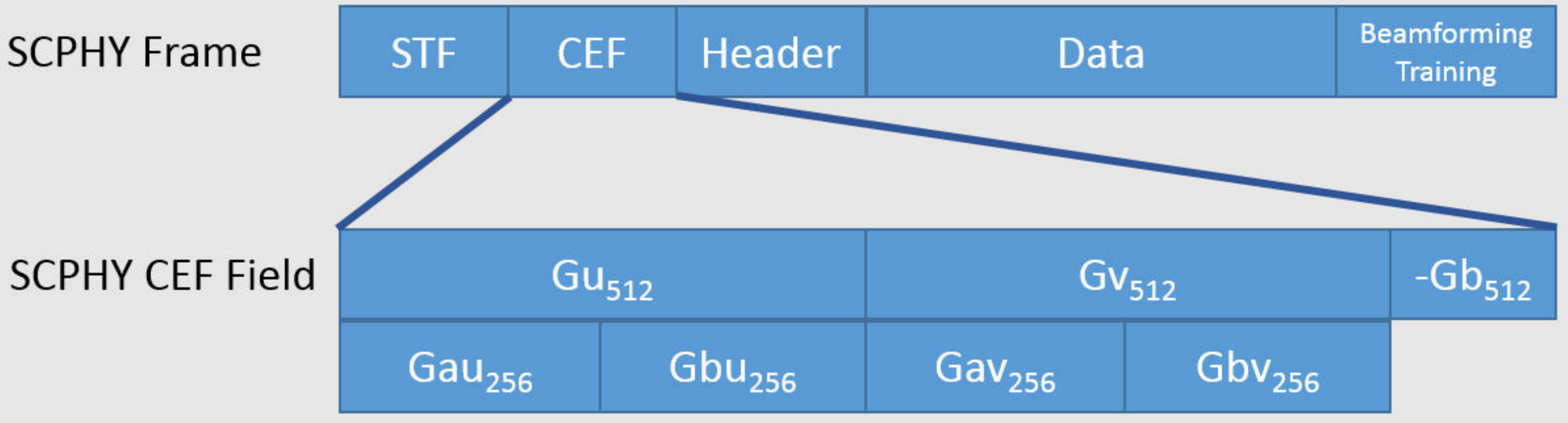}
  \caption{\hspace{-4pt}Composition of IEEE 802.11ad SCPHY frame.\vspace{-14pt}}
	\label{fig:scphy}
\end{figure}

The IEEE 802.11ad SCPHY frame consists of a short training field (STF), a channel estimation field (CEF), header, data and beamforming training field (Fig.~\ref{fig:scphy}). The STF and CEF together form the SCPHY preamble. Within the CEF lie two 512-point sequences $Gu_{512}[n]$ and $Gv_{512}[n]$. Each of them contains a Golay complementary pair of length $256$, $\{Gau_{256}, Gbu_{256}\}$ and $\{Gav_{256}, Gbv_{256}\}$, respectively.

A \textit{Golay complementary pair} consists of two sequences $Ga_N$ and $Gb_N$ each of the same length $N$ with entries $\pm1$, such that their \textit{aperiodic} autocorrelation functions have sidelobes equal in magnitude but opposite in sign. The sum of the two autocorrelations, therefore, has a peak of $2N$ and a sidelobe level of zero:\par\noindent\small
\begin{align}
\label{eq:golaytimeprop1}
Ga_N[n]*Ga_N[-n] + Gb_N[n]*Gb_N[-n] = 2N\delta[n],
\end{align}\normalsize
where $*$ denotes linear convolution. To observe this property in the frequency domain, consider $Fa_N$ and $Fb_N$ as the Discrete Fourier Transforms (DFTs) of the sequences $Ga_N$ and $Gb_N$, respectively. Then, the sum of the spectral densities of the two sequences is flat, i.e., for each DFT index $k$,\par\noindent\small
\begin{align}
\label{eq:golayfreqprop1}
|Fa_N[k]|^2 + |Fb_N[k]|^2 = 2N.
\end{align}\normalsize
The convolution in (\ref{eq:golaytimeprop1}) is linear while the DFT has the circular convolution property. Therefore, for these relations to hold, we must choose a DFT of length $N_{DFT} \ge 2N-1$. Other perfect autocorrelation sequences \cite{luke1988sequences} and complementary pairs \cite{hollis1962another} also exist and have been used for radar and communication applications. 

\begin{figure*}[ht]
	\centering
		\includegraphics[width=1.0\textwidth]{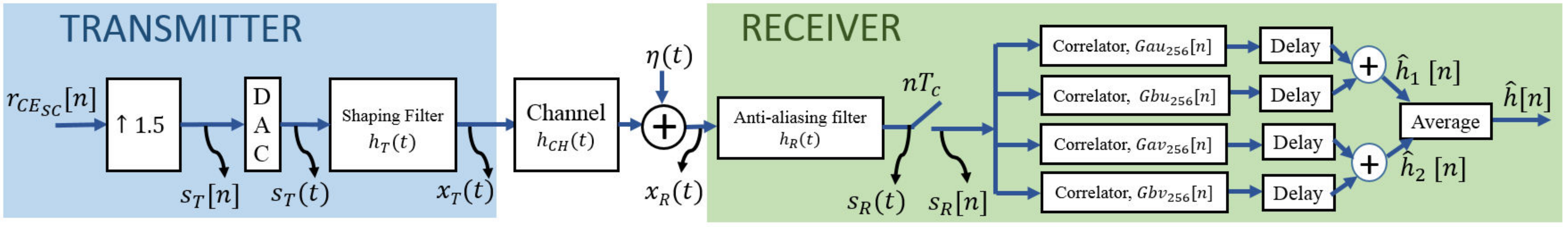}
		\caption{Nyquist processing for 802.11ad-based channel estimation.\vspace{-18pt}}
		\label{fig:NyquistFlow}
\end{figure*}
The 802.11ad CEF transmit signal is a concatenated sequence\par\noindent\small
\begin{align}
\label{eq:txseq}
r_{{{CE}_{SC}}}[n] &= Gu_{512}[n] + Gv_{512}[n-512], \phantom{1}n=0, 1, \cdots, 1151.
\end{align}\normalsize
Here, the sequences $Gu_{512}[n]$ and $Gv_{512}[n]$ are defined for $0 \le n \le 511$; for other values of $n$, they are set to zero. Typical 802.11ad transmitters employ DACs that operate at 2.64 GHz to also enable OFDM signals. Therefore, the discrete-time signal $r_{{CE_{SC}}}[n]$ is upsampled by a factor of $1.5$ to obtain a sequence $s_T[n]$ just before sending it to a digital-to-analog-converter (DAC) module (Fig.~\ref{fig:NyquistFlow}):\par\noindent\small
\begin{align}
r_{1_{CE_{SC}}}[n] &=\begin{dcases}
r_{{CE_{SC}}}[m],\: n=3m,\\
0, \:\mbox{otherwise},
\end{dcases}\\
s_T[n] &= r_{1_{CE_{SC}}}[2n].
\end{align}\normalsize
The corresponding upsampled constituting sequences of $s_T[n]$ are denoted as $Gu_{512_{up}}[n]$, $Gv_{512_{up}}[n]$, and so on. We represent the discrete-time sequence $s_T[n]$ as a weighted sum of Dirac impulses\par\noindent\small
\begin{align}
s_T(t) &= \sum\limits_{m=0}^{1535}s_T[m]\delta(t-mT_c),
\end{align}\normalsize
where $F_c = 2.64 GHz = 1/T_c$. This signal is passed through the transmit shaping filter $h_T(t)$ (Fig.~\ref{fig:NyquistFlow}) to obtain:\par\noindent\small
\begin{align}
x_T(t) = (s_T * h_T)(t).
\end{align}\normalsize
The 802.11ad protocol specifies a spectral mask for the transmit signal to limit inter-symbol interference \cite[section 21.3.2]{ieee2012phy80211ad}. We assume that $h_T(t)$ includes a low-pass baseband filter with an equivalent amplitude characteristic of the spectral mask.
\vspace{-8pt}
\subsection{Channel and receiver}
\label{subsec:nyqrx}
For the entire duration of the SCPHY packet, the channel is treated as linear and time-invariant \cite{rappaport2014millimeter}. Assuming the received signal contains echoes from $L$ paths, the continuous channel response is\par\noindent\small
\begin{align}
\label{eq:channel}
h_{CH}(t) = \sum_{l=1}^{L}\alpha_l.\delta(t-\tau_l),
\end{align}\normalsize
where $\alpha_l \in \mathbb{C}$ and $\tau_l$ are the path losses and delays, respectively. The received signal is then given by\par\noindent\small
\begin{align}
\label{eq:rxsig2}
x_R(t) = (x_T * h_{CH})(t) + \eta(t) = (s_T * h_T * h_{CH})(t) + \eta(t),
\end{align}\normalsize
where $\eta(t)$ is additive white Gaussian noise (AWGN). Prior to sampling at the receiver, an anti-aliasing continuous-time filter $h_R(t)$ (usually a milled analog filter at 60 GHz) is used to remove high-frequency noise from the signal $x_R(t)$ and also to attenuate adjacent channels. The signal processed by the filter $h_R(t)$ is\par\noindent\small
\begin{align}
\label{eq:rxsig1}
s_R(t) &= (x_R*h_R)(t) = (s_T * h_T * h_{CH}*h_R)(t) + (\eta*h_R)(t)\nonumber\\
&= (s_T* h_{TR}*h_{CH})(t) + z(t),
\end{align}\normalsize
where $h_{TR}(t) =  (h_T *h_R)(t)$ represents the overall analog filter impulse response and $z(t) = (\eta*h_R)(t)$ is the filtered noise. The received signal $s_R(t)$ is sampled at rate $F_c = 2.64$ GHz to obtain the discrete-time signal $s_R[n] = s_R(nT_c)$.
\vspace{-8pt}
\subsection{Channel estimation}
\label{subsec:nyqchanest}
In a Nyquist receiver, the received signal samples are downsampled by a factor of 1.5. Since the constant spectrum property (\ref{eq:golayfreqprop1}) of Golay sequences is not affected due to upsampling, here we skip the downsampling stage. The samples $s_R[n]$ are passed through four correlators, each corresponding to one of the Golay sequences of length $256$. For example, if we assume that the overall analog filter has a flat unity frequency response over the band of interest, then the \textit{signal trail} of one of the correlators is\par\noindent\small
\begin{align}
\hat{h}_{1au}[n] &= s_R[n]*Gau_{256}[-n] = (h_{CH}(nT_c)*s_T[n])*Gau_{256}[-n].
\end{align}\normalsize
By correlating the received samples with $Gbu_{256}$, $Gav_{256}$, and $Gbv_{256}$, we similarly obtain $\hat{h}_{1bu}[n]$, $\hat{h}_{1av}[n]$, and $\hat{h}_{1bv}[n]$.

Since we intend to exploit the property in (\ref{eq:golaytimeprop1}), we delay the correlator outputs: $\hat{h}_{1a}[n] = \hat{h}_{1au}[n]$,$\hat{h}_{1b}[n] = \hat{h}_{1bu}[n+256]$, $\hat{h}_{2a}[n] = \hat{h}_{2av}[n+512]$, $\hat{h}_{2b}[n] = \hat{h}_{2bv}[n+768]$. The outputs of the complementary pairs are then summed up. As shown in Fig.~\ref{fig:NyquistFlow}, this results in two estimates of the channel, $\hat{h_1}[n]$ and $\hat{h_2}[n]$. For instance, the first estimate of the channel is\par\noindent\small
\begin{align}
\hat{h}_{1}[n] &= \hat{h}_{1a}[n] + \hat{h}_{1b}[n]\nonumber\\			   
               &= h_{CH}[n]*(Gau_{256}[n]*Gau_{256}[-n] + Gbu_{256}[n]*Gbu_{256}[-n])\nonumber\\
               &= 512\cdot h_{CH}[n].
\end{align}\normalsize
We average to obtain the final estimate $\hat{h}[n] = \frac{1}{512}\frac{1}{2}(\hat{h}_1[n] + \hat{h}_2[n])$. In the absence of noise, $\hat{h}[n]$ recovers the true channel perfectly.
\vspace{-8pt}
\section{Sub-Nyquist Channel Estimation}
\label{sec:subnyqch}
\begin{figure*}[ht]
	\centering
		\includegraphics[width=1.0\textwidth]{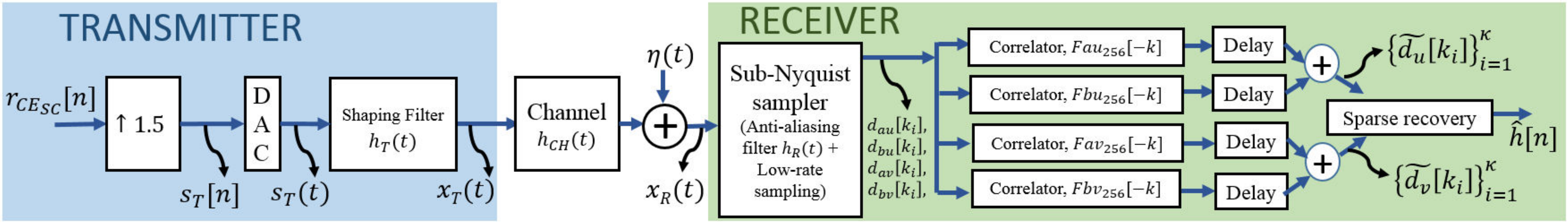}
		\caption{Processing flow for 802.11ad protocol channel estimation using sub-Nyquist sampling.\vspace{-13pt}}
		\label{fig:subNyquistFlow}
\end{figure*}
In order to reduce the GHz sampling rates of 802.11ad receivers, we propose sub-Nyquist channel estimation. It has been shown \cite{vetterli2002sampling,baransky2014sub,mishra2016cognitive} that sparse channels as in (\ref{eq:channel}) can be recovered from low-rate samples using only $K \ge 2L$ Fourier coefficients. We use the Xampling framework to obtain Fourier coefficients from low-rate samples. 
\vspace{-8pt}
\subsection{Fourier series representation}
\label{sec:fourier}
The sequence $s_T(t)$ consists of two Golay complementary pairs and each pair is processed separately. Therefore, hereafter, it suffices to consider only one pair (say $\{Gau_{256_{up}}(t), Gbu_{256_{up}}(t)\}$) for our discussion. The continuous-time signals $Gau_{256_{up}}(t)$ and $Gbu_{256_{up}}(t)$ are derived as weighted sums of Diracs from their discrete-time counterparts. The received signal corresponding to one of the transmit Golay sequences $Gau_{256_{up}}(t)$ is \par\indent\small
\begin{align}
\label{eq:finrxsig2}
s_{au}(t) &= (Gau_{256_{up}}* h_{TR}*h_{CH})(t) + z(t).
\end{align}\normalsize
The channel impulse response is of finite duration and we observe the signal for time $\tau_{CH}$. The Fourier series representation of the received signal in (\ref{eq:finrxsig2}) is\par\indent\small
\begin{align}
\label{eq:rxsig4}
s_{au}(t) = \sum_{k\in \mathbb{Z}}d_{au}[k]e^{j2\pi k t/\tau_{CH}},\phantom{1} t \in [0, \tau_{CH}].
\end{align}\normalsize
Here, $d_{au}[k]$ are the Fourier series coefficients:\par\indent\small
\begin{align}
\label{eq:signalnoisecoeff1}
d_{au}[k] = c_{au}[k] + Z[k],
\end{align}\normalsize
where $c_{au}[k]$ and $Z[k]$ are the Fourier series coefficients of the signal and noise trails. In particular,\par\indent\small
\begin{align}
c_{au}[k] &= \frac{1}{\tau_{CH}} \int\limits_{0}^{\tau_{CH}}s_{au}(t)e^{-j2 \pi k t /\tau_{CH}} dt\nonumber\\
&= \frac{1}{\tau_{CH}} \int\limits_{0}^{\tau_{CH}}(\underbrace{Gau_{256_{up}}*h_{TR} }_{=r_{au}(t)}*h_{CH})(t)e^{-j2 \pi k t /\tau_{CH}} dt\nonumber\\
&= \frac{1}{\tau_{CH}} \sum\limits_{l=0}^{L-1}\alpha_l\int\limits_{0}^{\tau_{CH}}r_{au}(t-\tau_l)e^{-j2 \pi k (t-\tau_l) /\tau_{CH}}e^{-j2 \pi k\tau_l /\tau_{CH}} dt.\label{eq:fourier}
\end{align}\normalsize

Suppose the continuous-time Fourier Transforms (CTFTs) of $Gau_{512_{up}}(t)$ and $h_{TR}(t)$ are $F_{au}(\omega)$ and $H_{TR}(\omega)$, respectively. Then, the CTFT of $r_{au}(t)$ is $R_{au}(\omega) = \int\limits_{-\infty}^{\infty}r_{au}(t)e^{-j\omega t} dt =  F_{au}(\omega)H_{TR}(\omega)$. Substituting into (\ref{eq:fourier}), we get\par\noindent\small
\begin{align}
c_{au}[k] &= \frac{1}{\tau_{CH}} R_{au}(2\pi k/\tau_{CH}) \sum\limits_{l=0}^{L-1}\alpha_le^{-j2 \pi k\tau_l /\tau_{CH}} \nonumber\\
&= \frac{1}{\tau_{CH}} F_{au}(2\pi k/\tau_{CH})H_{TR}(2\pi k/\tau_{CH}) \sum\limits_{l=0}^{L-1}\alpha_le^{-j2 \pi k\tau_l /\tau_{CH}} \label{eq:signalcoeff1}.
\end{align}\normalsize
If the low-pass filter $h_{TR}(t)$ has a cut-off frequency $f_{T}$ (say), then the Fourier series coefficients of the filtered signal above $f_{T}$ vanish:\par\noindent\small
\begin{align}
c_{au}[k] = 0,\: \forall \: k \ge \frac{f_{T}\tau_{CH}}{2\pi}.
\end{align}\normalsize
Following the aforementioned procedure, we similarly obtain $d_{bu}[k]$, $d_{av}[k]$, and $d_{bv}[k]$ corresponding to continuous-time Golay sequence signals $Gbu_{256}(t)$, $Gav_{256}(t)$, and $Gbv_{256}(t)$, respectively. 
\vspace{-8pt}
\subsection{Sampling and matched filtering} 
\label{subsec:lowrate}
Substituting (\ref{eq:signalcoeff1}) into (\ref{eq:signalnoisecoeff1}) produces\par\noindent\small
\begin{align}
\label{eq:signalnoisecoeff2}
d_{au}[k] = \frac{1}{\tau_{CH}} F_{au}(2\pi k/\tau_{CH})H_{TR}(2\pi k/\tau_{CH}) \sum\limits_{l=0}^{L-1}\alpha_le^{-j2 \pi k\tau_l /\tau_{CH}} + Z[k].
\end{align}\normalsize
The time delays are aligned to a grid $\tau_l = n_l\Delta_\tau$ where $0\le n_l \le N$ and $N = \tau_{CH}/\Delta_\tau$ is an integer. In Nyquist processing, all $N$ complex Fourier coefficients are obtained. For sub-Nyquist processing, the received signal is sampled at low rates. Corresponding to these low-rate samples, let $\kappa$ denote a subset of $N$ complex Fourier coefficients such that $\kappa = \{k_1, \cdots, k_{|\kappa|} \}$ and $k_{|\kappa|} < \frac{f_T\tau_{CH}}{2\pi}$. In Fig.~\ref{fig:subNyquistFlow}, this process corresponds to the ``Sub-Nyquist sampler'' block and yields $|\kappa| = K$ Fourier coefficients.

When the sampled signal is correlated with the Golay sequence $Gau_{256_{up}}$, the Fourier coefficients of the resulting signal are\par\noindent\small
\begin{align}
\tilde{d}_{au}[k] &= d_{au}[k]F_{au}(-2\pi k/\tau_{CH})\nonumber\\
&= \frac{1}{\tau_{CH}} |F_{au}(2\pi k/\tau_{CH})|^2 H_{TR}(2\pi k/\tau_{CH})\sum\limits_{l=0}^{L-1}\alpha_le^{-j2 \pi k\tau_l /\tau_{CH}}\nonumber\\
&\phantom{1}\phantom{1}+ Z[k]F_{au}(-2\pi k/\tau_{CH}),
\end{align}\normalsize
for all $k \in |\kappa|$. Similarly, for the other part of the complementary pair,\par\noindent\small
\begin{align}
\tilde{d}_{bu}[k] &= \frac{1}{\tau_{CH}} |F_{bu}(2\pi k/\tau_{CH})|^2 H_{TR}(2\pi k/\tau_{CH})\sum\limits_{l=0}^{L-1}\alpha_le^{-j2 \pi k\tau_l /\tau_{CH}}\nonumber\\
&\phantom{1}\phantom{1}+ Z[k]F_{bu}(-2\pi k/\tau_{CH}).
\end{align}\normalsize
The sum of the two complementary Fourier coefficients yields\par\noindent\small
\begin{flalign}
\tilde{d}_{u}[k] &= \tilde{d}_{au}[k] + \tilde{d}_{bu}[k]\nonumber\\
&= \frac{1}{\tau_{CH}}\underbrace{\{|F_{au}(2\pi k/\tau_{CH})|^2 + |F_{bu}(2\pi k/\tau_{CH})|^2\}}_{C_u}H_{TR}(2\pi k/\tau_{CH})\nonumber\\
&\phantom{1}\phantom{1}\phantom{1}\phantom{1}\phantom{1}\cdot\sum\limits_{l=0}^{L-1}\alpha_le^{-j2 \pi k\tau_l /\tau_{CH}} + \tilde{Z}[k]\nonumber\\
&= \frac{C_u}{\tau_{CH}} H_{TR}(2\pi k/\tau_{CH})\sum\limits_{l=0}^{L-1}\alpha_le^{-j2 \pi k\tau_l /\tau_{CH}} + \tilde{Z}[k],\label{eq:coeffrc3}
\end{flalign}\normalsize
where $C_u = 512$ (i.e. twice the Golay sequence length), and $\tilde{Z}[k] = Z[k]F_{au}(-2\pi k/\tau_{CH}) + Z[k]F_{bu}(-2\pi k/\tau_{CH})$ is the corresponding noise trail. We assume that only AWGN enters the receiver so that the filtered, sampled noise is white discrete noise. If the power density of white noise $z(t)$ is $N_0/2$ and the bandwidth of the anti-aliasing filter $h_R(t)$ is $B$, then the power of discrete white noise is $N_0B/2$. For both Nyquist and sub-Nyquist processing, the response $H_{TR}(2\pi k/\tau_{CH})$ is typically known via prior calibration.
\vspace{-8pt}
\subsection{Sparse recovery}
\label{subsec:sparse}
We can express the coefficients in (\ref{eq:coeffrc3}) in matrix form as\par\noindent\small
\begin{align}
\label{eq:csrecover1}
\mathbf{\tilde{d}_{u}} &= \frac{C_u}{\tau_{CH}}\mathbf{H}_{TR}\mathbf{B}\mathbf{y_1} + \mathbf{\tilde{z}},
\end{align}\normalsize
where $\mathbf{\tilde{d}_{u}} = \begin{bmatrix}\tilde{d}_{u}[k_1] & \cdots & \tilde{d}_{u}[k_{|\kappa|}]\end{bmatrix}^T$, $\mathbf{y_1}$ is an $L$-sparse vector of length $N$ whose non-zero values are the path losses $\alpha_l$, $\mathbf{H}_{TR} = diag\left(H_{TR}(2\pi k_1/\tau_{CH}),\cdots,H_{TR}(2\pi k_{|\kappa|}/\tau_{CH})\right)$ is a $K \times K$ diagonal matrix, $\mathbf{\tilde{z}}$ is the noise vector, and $\mathbf{B}$ is a $K \times N$ matrix with $\{l,m\}$th element $\mathbf{B}_{lm}=e^{-j2\pi k_lm/N}$. Equation (\ref{eq:csrecover1}) models a sparse recovery problem \par\noindent\small
\begin{flalign}
\label{eq:csrecoverl1_1}
	& \underset{\mathbf{y_1}}{\text{minimize}}\phantom{1}\left\Vert \mathbf{y_1}\right\Vert _{0}\nonumber\\
	& \text{subject to}\phantom{1} \left\Vert\mathbf{\tilde{d}_{u}} - \frac{C_u}{\tau_{CH}}\mathbf{H_{TR}}\mathbf{B}\mathbf{y_1}\right\Vert_2 \le \sigma_1,
\end{flalign}\normalsize
where $\sigma_1$ is some positive constant and $\|\cdot\|_0$ is the number of non-zero elements of the vector. This problem can be solved using one of the several compressed sensing recovery algorithms such as orthogonal matching pursuit (OMP) and $\ell_1$ minimization \cite{eldar2012cs}. Let the solution of (\ref{eq:csrecoverl1_1}) be $\mathbf{\hat{y}}_1$. For the other Golay sequence pair, we similarly obtain $\mathbf{\tilde{d}_{v}}$ and the corresponding sparse vector $\mathbf{\hat{y}}_2$ for a different set of sampled coefficients. Our final channel estimate is then $\mathbf{\hat{y}} = (\mathbf{\hat{y}}_1 + \mathbf{\hat{y}}_2)/2$. 

Alternatively, if the sampling coefficients are the same for both Golay pairs, then we first average to obtain $\mathbf{\tilde{d}} = (\mathbf{\tilde{d}_{u}} + \mathbf{\tilde{d}_{v}})/2$ and then obtain the channel estimate $\mathbf{\hat{y}}$ by solving a single sparse recovery problem \par\noindent\small
\begin{flalign}
\label{eq:csrecoverl1_2}
	& \underset{\mathbf{y}}{\text{minimize}}\phantom{1}\left\Vert \mathbf{y}\right\Vert _{0}\nonumber\\
	& \text{subject to}\phantom{1} \left\Vert\mathbf{\tilde{d}} - \frac{C_u}{\tau_{CH}}\mathbf{H_{TR}}\mathbf{B}\mathbf{y}\right\Vert_2 \le \sigma,
\end{flalign}\normalsize
where $\sigma$ is some positive constant.
\vspace{-8pt}
\section{Sub-Nyquist Sampling Methods}
\label{sec:numexp}
\begin{figure*}[ht]
	\centering
		\includegraphics[width=0.96\textwidth]{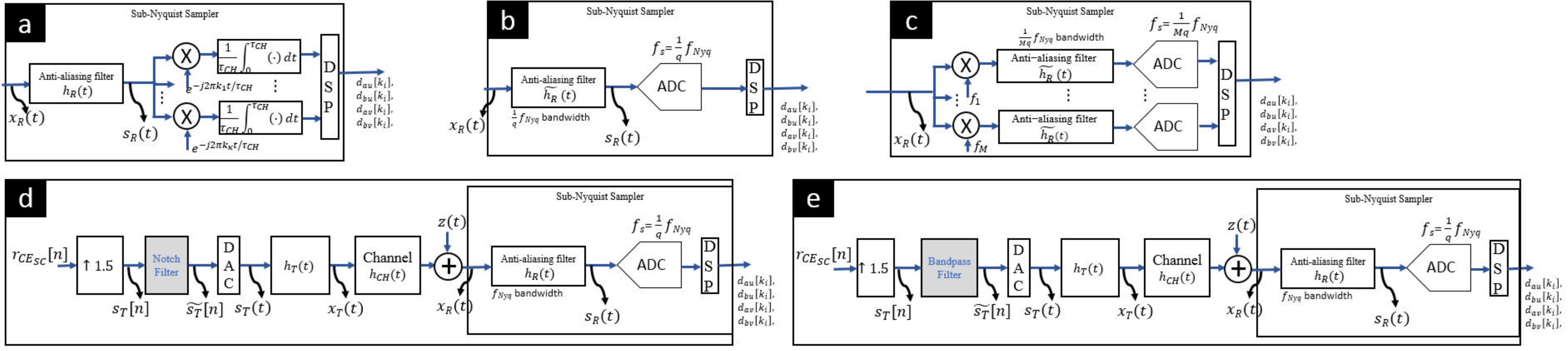}
		\caption{Sub-Nyquist sampling methods: (a) direct sampling (b) low frequencies only (c) multiband sampling (d) foldable sampling (e) foldable multiband.\vspace{-8pt}}
		\label{fig:sampling}
\end{figure*}
\begin{figure*}[ht]
	\centering
		\includegraphics[width=0.96\textwidth]{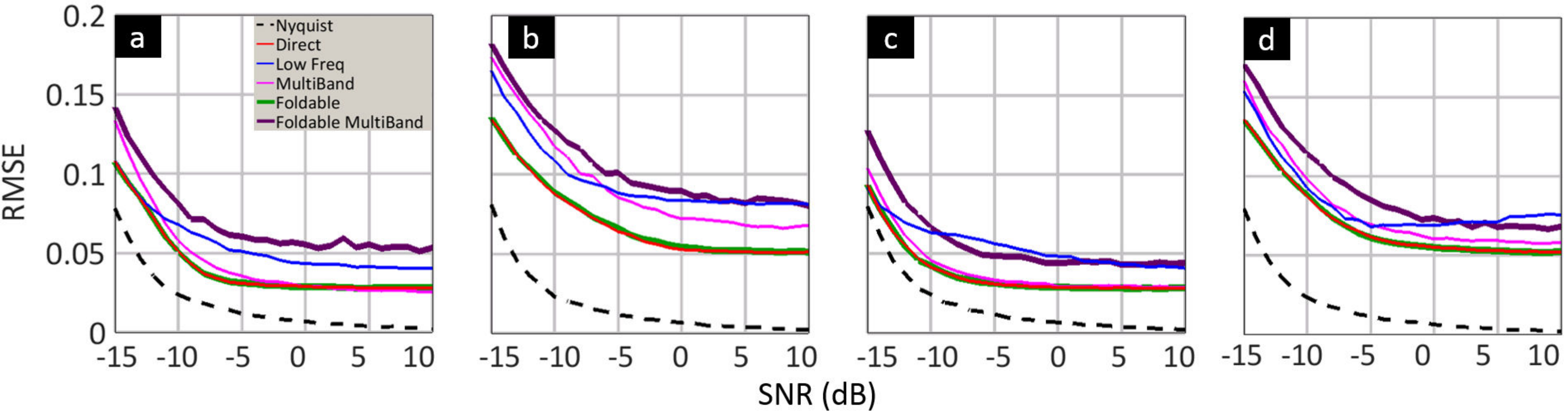}
		\caption{Performance of average-first-OMP-next algorithm with reduction factor (a) $q=2$ (b) $q=7$. OMP-first-average-next with (c) $q=2$ (d) $q=7$.\vspace{-16pt}}
		\label{fig:perf}
\end{figure*}
There are several ways to implement the sub-Nyquist sampler. For simplicity, consider $|\kappa| = K$ such that $q=N/K$ is an integer defining the sampling reduction factor. The 802.11ad single carrier Nyquist rate is $f_{Nyq} = F_c = 1/T_c = 2.64$ GHz. In \textit{direct sampling} (Fig.~\ref{fig:sampling}a), the signal $s_R(t)$ obtained after the anti-aliasing filter $h_R(t)$ is passed through as many analog chains as the number of sub-Nyquist coefficients $K$. Each branch is modulated by a complex exponential, followed by integration over $\tau_{CH}$. This technique gives most flexibility in choosing the Fourier coefficients, but is expensive in terms of hardware.\vspace{-3pt}

Another approach is to limit the bandwidth of $h_R(t)$ such that only the \textit{lowest $K$} frequencies are free of aliasing (Fig.~\ref{fig:sampling}b). Then, we sample these lowest $K$ frequencies with a (Nyquist) low-rate ADC. This method requires modification in 60 GHz hardware so that the analog receive filter $\tilde{h}_R(t)$ has reduced passband. In \textit{multiband} sampling, $M$ disjoint randomly-chosen groups of consecutive Fourier coefficients are obtained such that the total sampled coefficients are still $K$. This translates to splitting the signal across $M$ branches, passing the downconverted signal through reduced-bandwidth anti-aliasing filters $\tilde{h}_R(t)$, and then performing Nyquist sampling on each band with a low-rate ADC (Fig.~\ref{fig:sampling}c).\vspace{-3pt} 

The previous three methods alter only the receiver. We now discuss two sampling procedures where both transmitter and receiver are modified. In \textit{foldable sampling} \cite{cohen2014channel}, the received signal is sampled below the Nyquist rate resulting in folding over or aliasing of the signal spectrum that exceeds the Nyquist sampling rate of the ADC. Aliasing is avoided by notching out those coefficients from the transmit signal that will alias over the low-frequency part of the sampled signal. The entire signal consisting of $N$ DFT taps can be assumed to be divided into $q$ groups of $K$ coefficients. If we subsample this signal at the rate $f_{Nyq}/q$, then $K$ coefficients of each group will alias over each other in the sampled signal. But if $q-1$ coefficients at each tap are notched out in the transmit signal, then aliasing is circumvented. We randomly choose each of the $K$ coefficients from any of the $q$ groups and notch out the remaining $q-1$ coefficients in the transmit signal. The notch filter is a digital filter (Fig.~\ref{fig:sampling}d), and can be realized by zeroing out the undesired FFT coefficients of the signal $s_T[n]$. Let $\mathbf{1}_{\kappa}$ be a vector that is one only at coefficients in the set $\kappa$ and zero everywhere. Then, the transmit signal $\tilde{s_T}[n]$ is obtained by $\tilde{s_T}[n] = IFFT_{N}\{FFT_{N} \{s_T[n]\} \cdot \mathbf{1}_{\kappa} \}$, where $`\cdot'$ denotes entry-wise product. This has more flexibility in randomizing the selected coefficients over the full signal  spectrum. Changes in the 60 GHz filter $h_R(t)$ are also not necessary. This sampling may suffer from SNR loss due to aliasing of out-band noise; fortunately, the stopband attenuation of 60 GHz filters is sufficiently high to avert this. The \textit{foldable multiband} sampling (Fig.~\ref{fig:sampling}e), is similar to foldable sampling except that instead of selecting each coefficient randomly from $q$ groups, we choose $M$ sets of consecutive coefficients from each group such that the total number of coefficients is still $K$. The corresponding transmitter uses digital bandpass filters.\vspace{-3pt} 

We verified our procedures through numerical experiments. We used channel impulse responses of length $192$ with a sparsity level $K = 10$ and a flat response for the filter $H_{TR}$ as is typical in 802.11ad scenarios. We used four bands for multiband and multiband foldable sampling. The effective SNR for foldable methods after sampling is kept the same as the Nyquist in these experiments showing an improvement in RMSE at low SNRs; here, the noise in unsampled frequencies is completely filtered out. We characterized the recovery performance through the root mean square error: RMSE $= \sqrt{1/N\sum_{n=0}^{N-1}|h_{CH}[n] - \hat{y}[n]|^2}$. Fig.~\ref{fig:perf} shows the estimation performance for different algorithms, reduction factors (q = 2 and 7) and SNR levels. We observe that the foldable and direct sampling are closest to the Nyquist sampling in performance especially at low SNRs and high reduction factor $q$. At high SNRs, Nyquist sampling shows superior performance over all other methods. The performance of average-first-OMP-next is slightly better than averaging the individual optimized solutions.
\vspace{-16pt}
\section{Summary}
\label{sec:summary}
We presented a sub-Nyquist channel estimation procedure for the IEEE 802.11ad link. Among various low-rate sampling options, the foldable sampling method offers minimum degradation in RMSE and can be realized using a digitally programmable transmitter. Other methods necessitate changing the analog receive filter which is technologically difficult at 60 GHz. With some tolerance in performance loss, sub-Nyquist sampling methods can complement the promise of very high throughput in IEEE 802.11ad links.
\vspace{-8pt}
\section*{Acknowledgements}
\label{sec:ack}
We would like to thank Alecsander Eitan of Qualcomm Israel, Tamir Bendory and Kfir Cohen for  insightful conversations.

\newcommand{\BIBdecl}{\setlength{\itemsep}{0.001 em}}
\bibliographystyle{IEEEtran}
\scriptsize{\bibliography{main}}

\end{document}